\documentstyle[11pt,newpasp,twoside,epsf]{article}
\def\fa{{151 {\rm \thinspace MHz}}}
\def\fc{{1.4 {\rm \thinspace GHz}}}
\def\fd{{2.7 {\rm \thinspace GHz}}}
\def\fe{{5 {\rm \thinspace GHz}}}
\def\vvm{{$\langle V/V_{max} \rangle$ }}
\def\zmean{{$\langle z \rangle$ }}

\def\edcomment#1{\iffalse\marginpar{\raggedright\sl#1\/}\else\relax\fi}
\reversemarginpar

\begin{document}
 \title{Extragalactic radio source evolution \& unification: \\
clues to the demographics of blazars}
 \author{C A Jackson}
\affil{Research School of Astronomy \& Astrophysics,
The Australian National University, Cotter Road,  Weston Creek,
ACT 2611, Australia}
\author{J V Wall}
\affil{Department of Astrophysics, University of Oxford, Nuclear and
 Astrophysics Laboratory, Keble Road, Oxford OX1 3RH, UK}

\begin{abstract}

In this paper we discuss the demographics of the radio blazar
population: (i) what are their parent (`unbeamed') sources and (ii)
what magnitude and/or type of evolution have they undergone ? The
discussion is based on models of radio source evolution and beaming
based on a `dual population' unification paradigm. These models,
developed from radio blazar properties in bright samples, predict
blazar demographic trends at the lower flux-density levels; samples
from deep mJy-level surveys ({\it e.g.} NVSS and FIRST) may now
provide direct tests of these predictions.

\end{abstract}

\section{Dual-population Unification}

All galaxies are radio galaxies in a sense. The scale of activity
ranges from `normal' galaxies (like our own) whose radio emission
emanates from supernova remnants.  At higher radio powers are the
`starburst' galaxies, greatly-enhanced stellar activity resulting in
stronger radio emission. At similar radio powers are the Seyfert
galaxies with their small-scale radio core-jet structures and with
clear indication of AGN activity.  At the top end of the scale are the
powerful radio galaxies, together with BL\,Lacs and quasars, the
latter two being collectively termed blazars.

Radio galaxies tend to have steep radio spectra and therefore dominate
low radio-frequency ($\nu <$ 0.5 GHz) surveys. The structures are
predominantly double-lobed; Fanaroff \& Riley (1974) compared the
distances between central maxima of the radio structures to overall
size and discovered a dichotomy. Sources with regions of maximum
brightness separated by more than 0.5 times the overall source size
are edge-brightened (FRII), while when the regions of maximum
brightness lies within this limit the source has diffuse outer lobes
(FRI) (centrally concentrated); and the highest radio power objects
are {\it predominantly} FRIIs whilst those of lower radio power are
{\it usually} FRIs. The overlap between the classes at middling radio
powers is seen in the local radio luminosity function (Figure 1), and
the division is known to be a function of optical luminosity as well
as radio (Owen and Ledlow 1993).

\begin{figure}
\plotfiddle{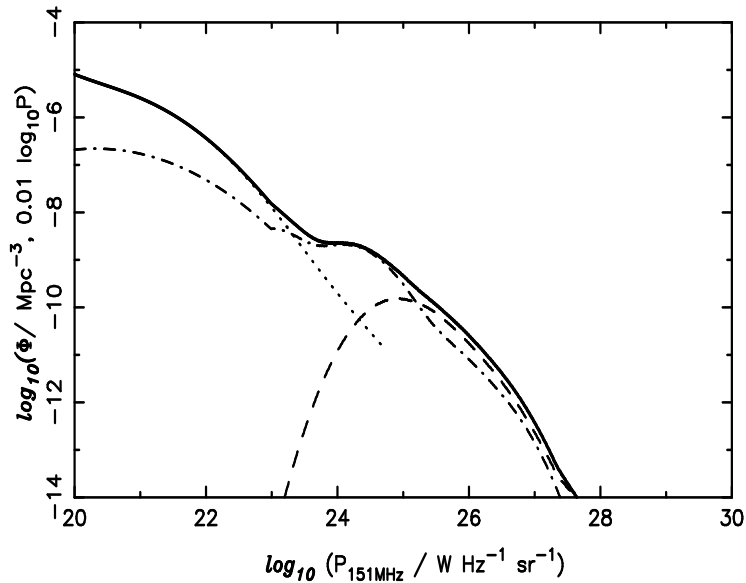}{3.0in}{270}{39}{45}{-160}{270}
\vspace*{-0.25in}\caption{The local radio luminosity 
function at 151 MHz (dark, solid line)
comprises contributions from
3 populations: starburst galaxies (dotted), FRI (dot-dash) 
and FRII radio galaxies (dashed).} 
\end{figure}

Evidence is accumulating that the powerful radio sources, double-lobed
radio galaxies on the one hand and blazars on the other, are `unified'
by projection effects, with a central torus together with relativistic
beaming responsible for our classification of objects as radio
galaxies or blazars (Scheuer \& Readhead 1979; Orr \& Browne 1982;
Scheuer 1987, Barthel 1989, Urry \& Padovani 1995 and references
therein). The `dual-population' unified scheme (Wall \& Jackson 1997),
summarised in Table 1, posits that FRII radio galaxies are the parent
population of {\it all} quasars and {\it some} BL\,Lac type objects.
The FRII-quasar sources are galaxies with class `A' spectra (Hine \&
Longair, 1979), having strong, high-excitation emission lines in their
optical/UV spectra. The FRII-BL\,Lac sources are galaxies with class
`B' spectra (Hine \& Longair, 1979), having only weak, if any,
high-excitation lines in their spectra.  The second part of the scheme
posits that FRI radio galaxies are the parent population of the
remainder of the BL\,Lac type objects.

\begin{table}
\caption{Unified classes of powerful extragalactic radio sources}
\label{hipops}
\vspace{0.1in}
\begin{tabular}{lccc}
          & UV/optical &  Radio \\
          & emission   &  spectrum & Source \\
Population & type      & at 5 GHz &  type \\

\hline
\\
FRII, class A & narrow & steep & RG \\
              & broad & steep & RG  \\
              & broad & `flat' & Quasar \\
\\
FRII, class B & weak/none & steep & RG \\
              & none & `flat' & BL Lac  \\
\\
FRI  & weak/none & steep & RG \\
              & none & `flat'  \\
\\
\hline
\end{tabular}
\end{table}

\section{The cosmic evolution of the parent populations}

Our analysis interprets radio source counts and and
identification/redshift data for samples at high flux densities,
adopting a dual-population unified scheme paradigm. Using a simple
parametric model to describe the evolution of the underlying parent
populations we first fit low-frequency radio data ($\nu <$ 0.5 GHz),
where radio samples are unbiased by the effects of Doppler beaming.
Then, using these evolution models for the parent populations of FRI
and FRII objects, we fit the 5 GHz radio source count using a small
number of parameters to describe the Doppler beaming, which gives rise
to blazars ({\it i.e.}  quasars and BL Lacs with $\alpha_{\fd}^{\fe} >
-0.5$, where $S \propto \nu^{\alpha}$)\footnote{Note that our
definition of blazar is solely a radio definition and says nothing
about the optical equivalent width criteria.}

The fundamental dataset to our analysis is the differential radio source
count, the count at a single observing frequency of source density on the sky as a function of flux density.   
Compiling data from wide-area and deep pencil-beam surveys 
yields source counts spanning 2 decades in frequency
and 6 decades in flux density as shown in Figure 2.

In fact radio source counts contain a wealth of cosmological
information, in particular providing potent evidence against a
steady-state Universe (Ryle \& Clarke 1961) and indicating that
differential evolution must take place, with the most luminous sources
showing the strongest evolution. In addition the {\it shape} of the
source count changes with observing frequency, suggesting that more
complex behaviour than just number density evolution is taking place -
and our analysis shows just what this behaviour is.

\begin{figure}
\plotfiddle{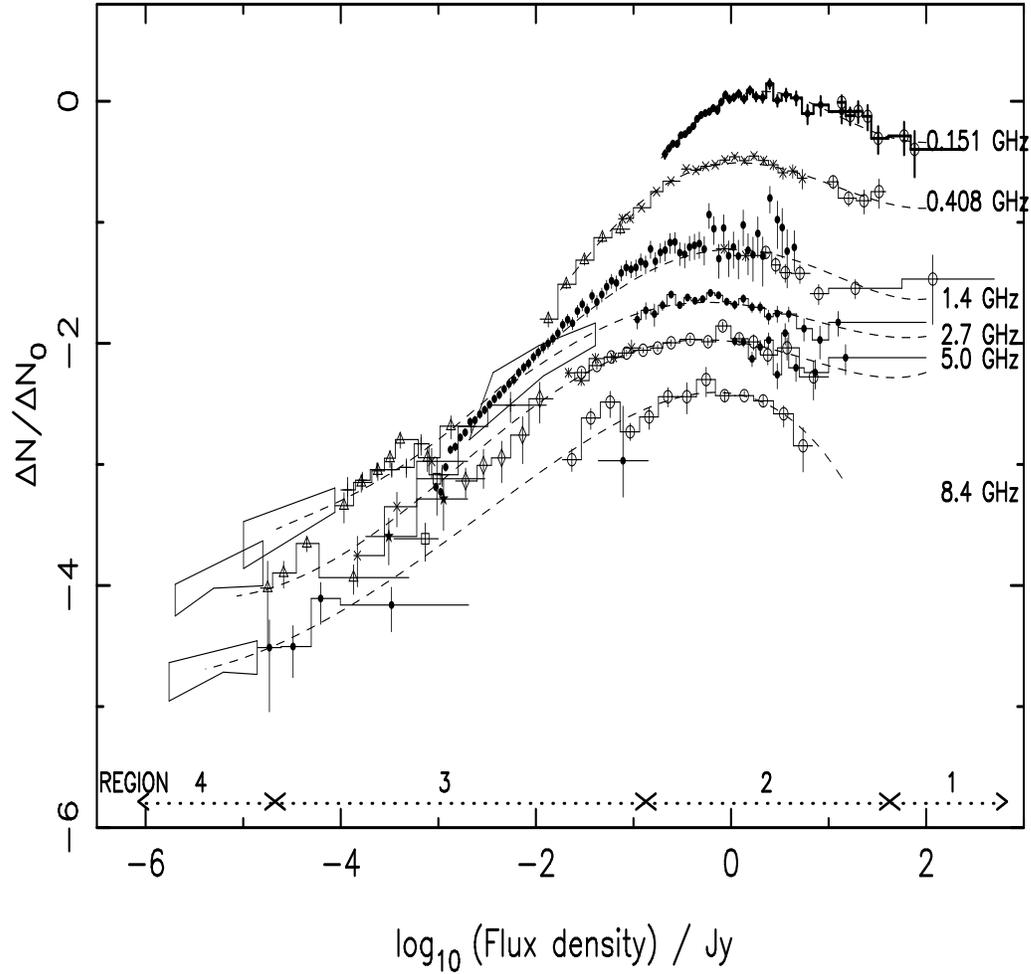}{4.0in}{270}{55}{78}{-220}{400}
\vspace*{0.6in}
\caption{Source counts in relative differential form $\Delta N/\Delta N_{0}$
where $\Delta N$
is the number of sources per sterad with flux density $S_{\nu}$ between
$S_{2}$ and $S_{1}$ and $\Delta N_{0}$ is the number
of sources expected in a uniformly-filled Euclidean universe
($N_{0} = K_{\nu} S_{\nu}^{-3/2}$
with $K_{\nu}$ = 2400, 2730, 3618, 4247, 5677 and 3738 for the six
frequencies shown). The horizontal bars show
the flux-density bin width, $S_{2}$ to $S_{1}$ and the vertical error
is the $\sqrt N$ error.
The curves are polynomial least-square fits to the counts.
The counts are compiled from radio survey data described in
Wall (1994).}
\end{figure}

Identification programmes and further studies of radio source 
samples have revealed distinct regions within the counts. 
These regions are indicated along the abscissa in Figure 2
and discussed briefly here:

\begin{description}

\item {\bf Region 1} At the very highest flux densities the source
count is near-Euclidean, although typically less than 20 sources
contribute to this region. This region is comprised of a mixture of
nearby sources and cosmologically-distant powerful sources. The flat
count arises due to the dilution of the powerful evolving sources by
the local lower-radio-power sources which are far more abundant at low
redshifts.

\item {\bf Region 2} In this region the counts rise more rapidly with
decreasing flux density than the Euclidean (-3/2 power law)
prediction, and then reach a Euclidean plateau.  This region of the
count is made up of the most powerful radio sources at cosmological
distances. The low-frequency counts ($<$ 0.5 GHz) comprise almost
entirely of steep-spectrum sources whose radio emission is greatest at
these frequencies.  The width of the plateau increases with increasing
frequency, due to an increasing blazar (flat-spectrum) contribution.

\item {\bf Region 3} Below the Euclidean plateau the counts at all
wavelengths drop away from the Euclidean prediction. This region
extends up to two orders of magnitude in flux density. (Limited)
identification data shows that the sources at such flux densities are
lower-power objects at intermediate redshifts and not, as might be
supposed, powerful radio sources at ever-increasing redshifts.  Very
little work has been done on the blazar population at these flux
density levels.

\item {\bf Region 4} At low flux densities the counts again flatten to
near-Euclidean. In this region the dominant populations change
dramatically to `blue' starburst galaxies and `red' FRI-type galaxies,
both relatively local and of low radio powers ({\it e.g.} Windhorst et
al. 1993, Benn et al. 1993).

\end{description}

Using the spectroscopically-complete 3CRR sample at 178 MHz 
(Laing, Riley \& Longair 1983 and more recently published
data collated by R. Laing, private communication) we first
determine an appropriate form for the evolution of the FRI
and FRII populations using the \vvm statistic. 

Figure 3 shows $V/V_{max}$ values for 137 steep-spectrum FRII radio
galaxies and 26 FRI radio galaxies.  Both populations show increasing
\vvm with radio power. However, FRIs exibit negative evolution with
\vvm $< 0.5$ at $\log_{10}(P_{\fa} \sim 25.0)$ and the FRIIs exhibit
strong positive evolution with \vvm $> 0.7$ at $\log_{10}(P_{\fa} \sim
27.5)$.  Interestingly, there is a trend for \vvm to continuously
increase with $\log_{10}(P_{\fa})$ across the populations, suggesting
luminosity-dependent evolution which may be population-independent.

\begin{figure}
\plotfiddle{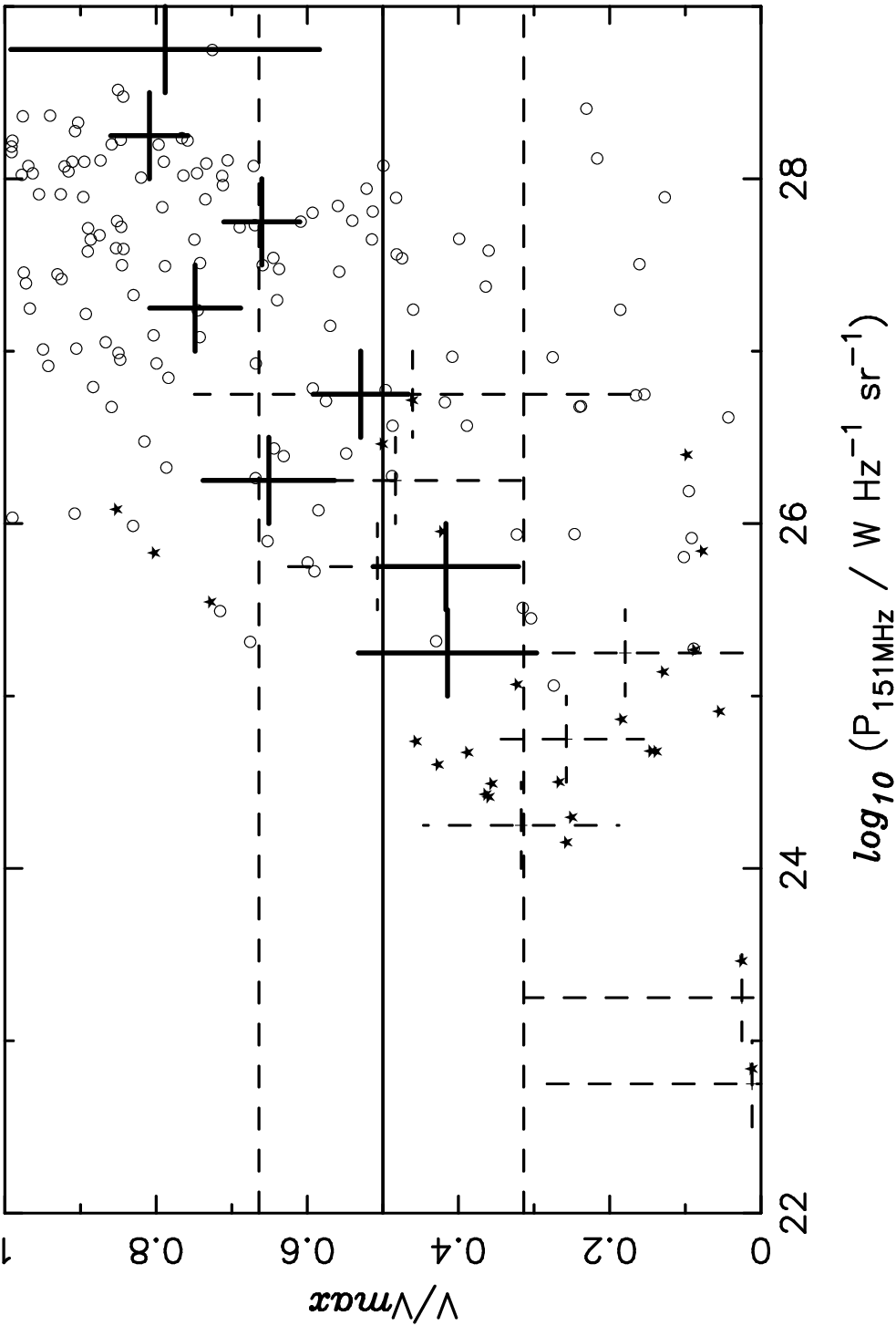}{3.0in}{270}{39}{45}{-205}{270}
\vspace*{-0.25in}\caption{$V/V_{max}$ for individual 3CRR FRI radio
galaxies ($\star$) and FRII radio galaxies ($\circ$). Overplotted are
\vvm values for bins of 0.5 in $\log_{10}(P_{\fa})$
for FRIs (dashed $+$) and FRIIs (solid $+$).} 
\vspace*{-205pt}\hspace*{272pt}\vvm for all FRIIs \\
\vspace*{50pt}\hspace*{272pt}\vvm for all FRIs
\vspace*{200pt}
\end{figure}

Adopting exponential luminosity-dependent density evolution (LDDE) we
fit the source count at 151 MHz comprising the 3CRR and 6C survey
(Hales, Baldwin, \& Warner 1988), using a minimisation routine.  The
model fit requires strong evolution of the most powerful FRII sources,
with little or no evolution of the FRI population (Table 2).  Thus the
fit reproduces the \vvm behaviour for the radio galaxy population as a
whole, in that only FRIIs with $\log_{10}(P_{\fa}) > 25.44$ undergo
any evolution, the lower-power FRIIs and all FRIs have a constant
space density to their cut-off redshift, $z_{c}$.

\begin{table}
\caption{Fitted evolution parameter values at 151 MHz \\
for LDDE:
$\rho(P,z) = \rho_{0}(P) \exp M(P) \tau(z)$}
\label{evolpops}
\vspace{0.1in}
\begin{tabular}{lcrr}
           & Exponential LDDE  & \multicolumn{2}{c}{Chi-square test} \\
Population & parameter values & $\chi^{2}_{min}$ & $\nu^{a}$ \\
\hline
\\
FRI  & $M_{max} = 0.0$, $z_{c}$ = 5.0 \\ 
     & $P_{1}$ \& $P_{2}$ not used given $M_{max}$ = 0.0 \\ 
\\
FRII, class A \& B & $M_{max} = 10.93$, $z_{c}$ = 5.62 \\ 
     & $P_{1} = 25.44$, $P_{2} = 27.34$ \\ 
\\
     & best-fit & 30.73 & 33 \\
\hline
\end{tabular}

$^{a}$ Degrees of freedom
\end{table}

\section{Blazars from beamed parent radio sources}

We fit the 5 GHz source count, which is well-defined over a wide
flux-density range, starting from the evolution model fit of Table 2.
However at 5 GHz we incorporate the beamed products (blazars) by
randomly aligning the sources with respect to our line-of-sight,
allowing both different Lorentz factors ($\gamma$) and different
core-to-extended flux ratios for the two (FRI,FRII) populations.  The
best-fit beaming parameters are shown in Table 3.

\begin{table}
\caption{Fitted beaming parameter values at 5 GHz}
\label{evolpops}
\vspace{0.1in}
\begin{tabular}{lcrr}
           &                  & \multicolumn{2}{c}{Chi-square test} \\
Population & Parameter values & $\chi^{2}_{min}$ & $\nu^{a}$ \\

\hline
\\
FRI  & $\gamma$ = 15.0, $R_{med} \propto P^{-0.55}_{\fa}$ \\
\\
FRII, class A \& B & $\gamma$ = 8.5, $R_{med}$ = 0.01 \\
     & $\theta_{c}(R_{med}) = 7^{\circ}.1$ \\
\\
     & best-fit & 32.98 & 25 \\
\hline
\end{tabular}
\end{table}

The contribution from each population to the 5-GHz source count
is shown in Figure 4. The contribution from the
quasar population peaks at a higher flux density limit 
than its parent population, and a similar behaviour is seen for the
the BL Lac objects. 

\begin{figure}
\plotfiddle{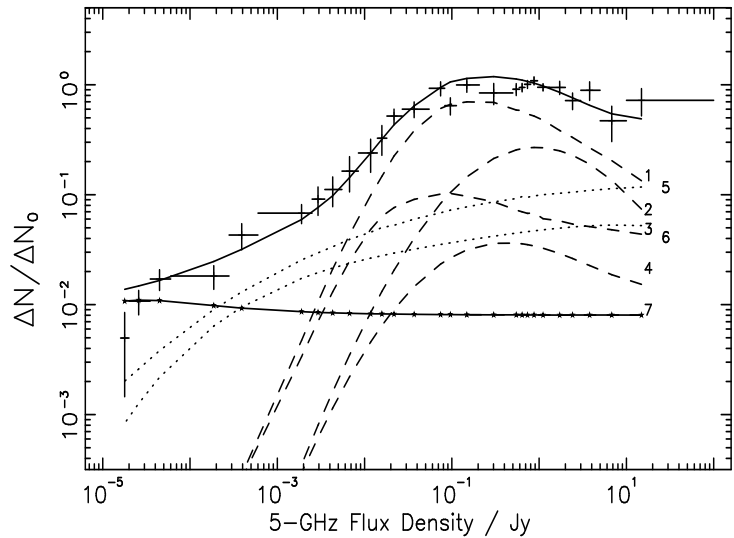}{3.0in}{270}{39}{45}{-160}{270}
\vspace*{-0.3in}\caption{Observed ({\large $+$}) and model 
(solid line)  differential source counts at 5 GHz. 
The model count comprises 4 contributions from FRII parents (dashed):
1) high-excitation radio galaxies, 2) quasars (from high-excitation
parents), 3) low-excitation radio galaxies and 4) BL Lac type
objects (from low-excitation parents). There are 2 contributions
from FRI parents (dotted): 5) radio galaxies and 6) BL Lac type
objects. The contribution from the starburst galaxy population 
is shown dot-dashed (7). }
\end{figure}

We can analyse the relative importance of each class of radio source
as a function of flux density. Blazars are 
dominated by quasars (high-excitation FRII parents) 
at $S_{\fe} >$ 0.3 Jy, but
dominated by BL Lac-type objects at lower flux densities. 
The magnitude of the rapidly changing quasar fraction predicted
by our model is well-matched by observations from the 0.25 Jy-sample
of flat-spectrum sources of Shaver et al. (1996) - see Figure 15 of
Jackson \& Wall (1999).

In terms of population mix, it is also instructive to consider the
integral population count, shown in Figure 5 in which the model fit
has been transposed to 1.4 GHz.  This figure reveals that complete
samples are predicted to comprise $\sim$20\% BL Lac objects for flux
density limits extending down to $S_{\fc} \sim$ 0.1 mJy.  Optical
spectroscopy of objects selected from the FIRST survey (Becker, White
\& Helfand 1995) could provide substantial tests of this prediction to
$S_{\fc} \sim$ 1 mJy.

\begin{figure}
\plotfiddle{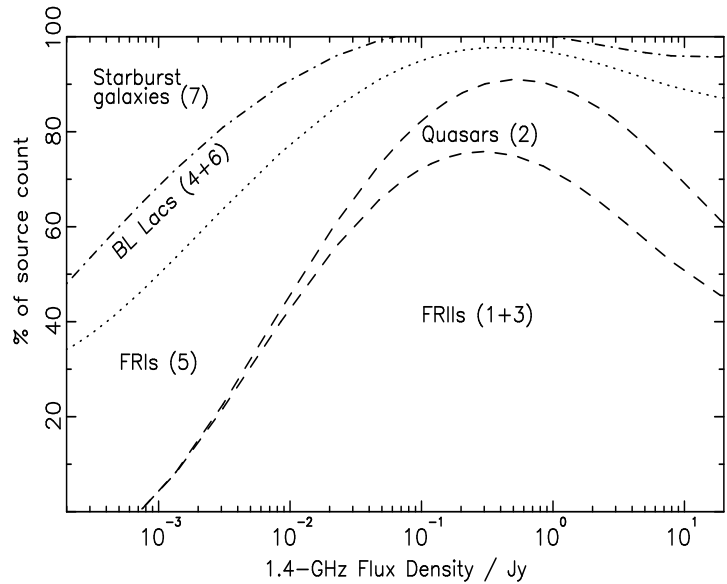}{3.0in}{270}{39}{45}{-160}{270}
\vspace*{-0.3in}\caption{Integral population mix 
from the model fit at 5 GHz transposed to 1.4 GHz. Numbers in
braces refer to the populations described in Figure 4. }
\end{figure}

\section{Redshift distributions of blazars}

Transposing our model (evolution plus beaming) to 2.7 GHz we 
compare our predictions with the limited 
observational data currently available.
Figure 6 shows the mean redshift, $\langle z \rangle$,  
across a wide flux density limit range for 
all source types as well as three sub-classes. 
For complete samples the model predicts that the value of \zmean peaks
at $\sim$20 mJy then drops rapidly. In detail, we find that
{\it (i)} the contribution from steep-spectrum sources dominates
and therefore traces the overall $\langle z \rangle$,
{\it (ii)} for the quasar population, \zmean rises continuously 
as $S_{\fd}$  decreases. However the {\it number} of 
quasars is negligible below
$S_{\fd}$ = 20 mJy so this has little effect on the
overall \zmean value and {\it (iii)}
the peak \zmean of the BL Lac sources occurs at a higher
flux density than that of the quasars, 
although the value of \zmean is much lower. 

Comparing the model \zmean values to the 0.25 Jy flat-spectrum 
quasar sample of Shaver et al. (1996 and in preparation) we find
reasonable agreement.
The data points for the 0.25 Jy sample shown
on Figure 6 are {\it lower} limits as this sample 
selects quasars with
($\alpha_{\fd}^{\fe} > -0.4$). Our model, however, 
defines blazars as having
($\alpha_{\fd}^{\fe} > -0.5$). Beamed objects with
$\alpha_{\fd}^{\fe} > -0.4$ 
have parent sources of lower radio luminosity, which in turn are 
predicted to have a lower \zmean
distribution than the less-beamed
$-0.5 < \alpha_{\fd}^{\fe} < -0.4$, higher-luminosity
objects which are missing from the 0.25~Jy sample.

The \zmean value for the complete 2-Jy sample (Wall \& Peacock 1985,
Morganti et al. 1997) agrees well with the overall \zmean model
prediction.

Figure 7 splits the BL\,Lac \zmean distribution 
of Figure 6 into its contributing
parent populations. We see that low-excitation-FRII BL\,Lac
sources have a \zmean distribution which 
is very similar to the quasar distribution. However,
the FRI BL\,Lacs have a very different distribution, with
\zmean gradually increasing with decreasing flux density.
Morphological studies of a sizeable sample of BL\,Lac objects
from {\it e.g.} the FIRST survey would provide direct, powerful
tests of this prediction.

\begin{figure}
\plotfiddle{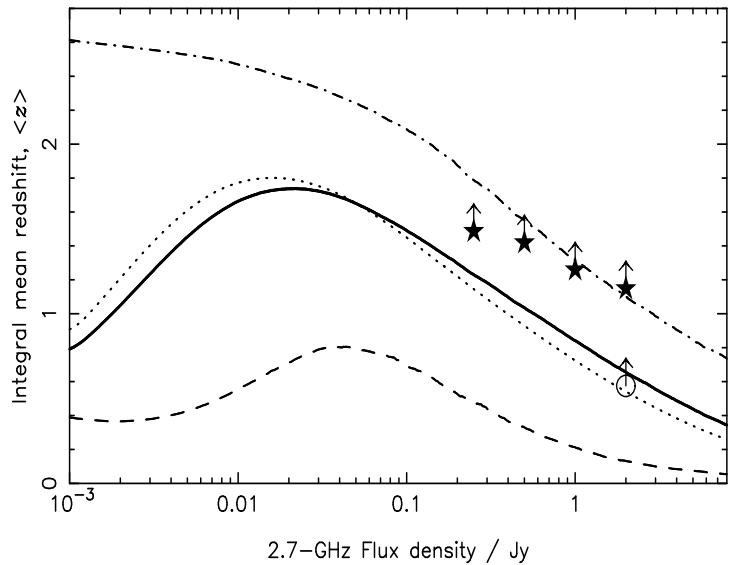}{3.0in}{270}{39}{45}{-160}{270}
\vspace*{-0.3in}\caption{Model \zmean of all 
sources to limiting flux density at 1.4 GHz (solid line). Contributions from 
the steep spectrum FRII, FRI and starburst galaxies (populations
1, 3, 5, and 7) shown dotted, 
quasars (from high-excitation FRIIs, population 2) dot-dashed and
BL Lacs (from low-excitation FRIIs and FRIs, populations 4 and 6)
dashed. Data points are from the 
flat-spectrum quasar sample ($\star$) (Shaver et al. 1996) 
and the 2 Jy sample ($\circ$) 
(Wall \& Peacock 1985, Morganti et al. 1997). 
The solid curve shows the total.}
\end{figure}

\begin{figure}
\plotfiddle{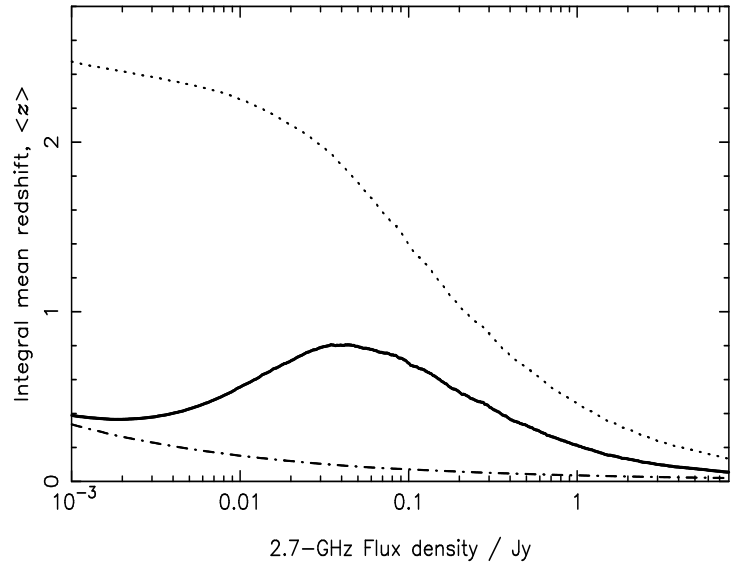}{3.0in}{270}{39}{45}{-160}{270}
\vspace*{-0.3in}\caption{Model mean redshift of 
BL Lacs from their two parent populations -- low-excitation 
FRIIs (population 4) dotted, FRIs (population 6) dot-dashed. }
\end{figure}


\begin{references}
\reference Barthel, P. D. 1989, ApJ, 336, 606 

\reference Benn, C. R., Rowan-Robinson, M., McMahon, R. G., 
Broadhurst, T. J., \& Lawrence, A. 1993, MNRAS, 263, 98

\reference Fanaroff, B. L., \& Riley, J. M. 1974, MNRAS, 167, 31P

\reference Hales, S. E. G., Baldwin, J. E.,  \& Warner, P. J. 1988, MNRAS
234, 919

\reference Hine, R. G., \& Longair, M. S. 1979, MNRAS, 188, 111

\reference Jackson, C. A., \& Wall, J. V. 1999, MNRAS, 304, 160

\reference Laing, R. A., Riley, J. M., \& Longair, M. S. 1983, MNRAS, 204, 151

\reference Longair, M. S. 1966, MNRAS, 133, 421

\reference Morganti, R., Oosterloo, T. A., Reynolds, J. E., 
Tadhunter, C. N., \& Migenes, V. 1997, MNRAS, 284, 541

\reference Orr, M. J. L. \& Browne, I. W. A. 1982, MNRAS, 200, 1067

\reference Owen, F. N. \& Ledlow, M. J. 1994, in The Physics of 
Active Galaxies, eds Bicknell, G. V. et al., ASP Conf. Ser., 54, 319

\reference Ryle, M., \& Clarke, R. W. 1961, MNRAS 122, 349
\reference Scheuer, P. A. G. 1987, in Superluminal Radio Sources, ed. 
J. A. Zensus \& T. J. Pearson, CUP, Cambridge, 104

\reference Scheuer, P. A. G., \& Readhead, A. C. S. 1979, Nature,
277,182

\reference Shaver, P. A., Wall, J. V., Kellermann, K. I., Jackson,
C. A., \& Hawkins, M. R. S. 1996, Nature, 384, 439

\reference Urry, C. M. \& Padovani, P. 1995, PASP, 107, 803

\reference Wall, J. V., \& Jackson, C. A. 1997,  MNRAS, 290, L17
\reference Wall, J. V., \&  Peacock, J. A. 1985, MNRAS, 216, 173

\reference Windhorst, R. A., Fomalont, E. B., Partridge, R. B., 
\& Lowenthal, J. D. 1993, ApJ, 405, 494


\end{references}
\end{document}